\documentclass[journal=jctcce,layout=twocolumn,manuscript=article]{achemso}

\usepackage{txfonts}
\usepackage[colorlinks=true, linkcolor=blue, urlcolor=blue, citecolor=blue]{hyperref}

\makeatletter
\let\l@addto@macro\relax
\makeatother
\usepackage[fontsize=11pt]{scrextend}

\title{QC Lab:\\A Python Package for Quantum--Classical Dynamics}

\author{Alex Krotz}
\affiliation{Department of Chemistry, Northwestern University, 2145 Sheridan Road, Evanston, Illinois 60208, USA}
\author{Ethan Byrd}
\affiliation{Department of Chemistry, Northwestern University, 2145 Sheridan Road, Evanston, Illinois 60208, USA}
\author{Ken Miyazaki}
\affiliation{Department of Chemistry, Northwestern University, 2145 Sheridan Road, Evanston, Illinois 60208, USA}
\author{Roel Tempelaar}
\affiliation{Department of Chemistry, Northwestern University, 2145 Sheridan Road, Evanston, Illinois 60208, USA}
\email{roel.tempelaar@northwestern.edu}

\begin{tocentry}
\includegraphics[width=\columnwidth]{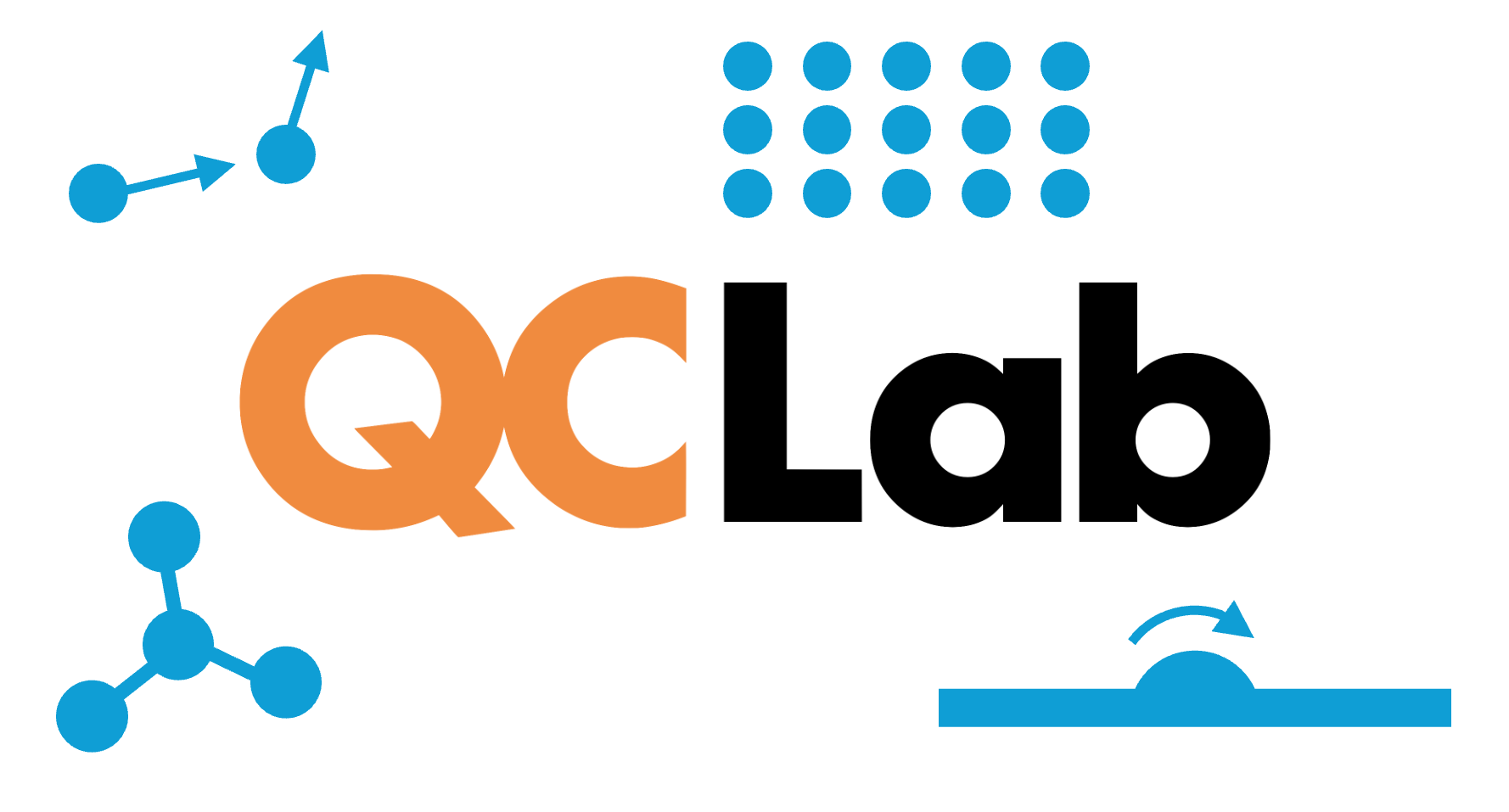}
\end{tocentry}

\begin{document}

\maketitle

\begin{abstract}
QC Lab is an open-source Python package for QC dynamics simulations aimed to promote the development of QC algorithms, and their application to a wide variety of relevant model problems. It follows a modular design that facilitates cross-compatibility between algorithms and models. By decomposing algorithms and models into a series of tasks and ingredients that can be substituted and reused, it minimizes development efforts and code redundancy. In this Paper, we introduce the first stable release of QC Lab, and describe its design philosophy.
\end{abstract}

\section{Introduction}

Many of the questions defining the frontier of chemistry, physics, and materials science involve the excited-state dynamics of an intractable quantum system.  The rationalization of such dynamics is increasingly reliant on computational algorithms that represent the dynamics as a quantum system of reduced complexity interacting with a surrounding ``bath'' that accounts for the surplus of involved coordinates in an approximate fashion.  Quantum--classical (QC) dynamics refers to the general approach where the bath coordinates are treated by inexpensive classical equations of motion.  Through the last few decades, QC dynamics has increasingly positioned itself at the forefront of excited-state dynamics research \cite{nelson_nonadiabatic_2014, subotnik_understanding_2016, wang_recent_2016, crespo-otero_recent_2018}, which is plausibly attributable to a variety of distinct advantages it provides.

First, and perhaps foremost, QC dynamics offers a level of efficiency and scalability that surpasses that of alternative approaches. It allows mixed states and ensemble effects to be incorporated by averaging over trajectories, each of which involves pure quantum states representable as state vectors. Although this introduces the need for trajectory averaging, it circumvents the density matrix representation that underpins quantum master equations; something that may prove prohibitively-expensive for sizable quantum systems. Second, QC dynamics treats the system--bath coupling non-perturbatively, permitting the application of QC methods to problems where such coupling is strong. (It should be noted that QC dynamics is not generally free from perturbative approximations, a topic that has been discussed in the literature \cite{herman_semiclasical_1984, kay_hermankluk_2006}.) Third, QC dynamics retains microscopic detail of the bath, allowing functional behaviors of the underlying coordinates to be uncovered. Additionally, there is a cultural/sociological advantage to QC dynamics, namely that it bridges between two major areas of computational chemistry, that is, electronic structure theory and molecular dynamics. As such, efforts in QC dynamics can readily build on decades worth of breakthroughs in both areas.

In spite of its relatively-low computational cost, QC dynamics remains expensive compared to static calculations, as a result of which an atomistic application is oftentimes not feasible. In such cases, it demands a careful construction of a model problem that faithfully captures the phenomenon of interest within a reduced representation. This may involve coarse graining, basis transformations, or creative use of the system--bath separation. Further, the adoption of the classical approximation exclusively for the bath coordinates presents a source of arbitrariness, giving rise to myriad ways in which QC algorithms can be constructed, each with their own realm of applicability. Here, a parallel can be drawn with the coexistence of myriad functionals in density functional theory. Importantly, functionals share a common structure, which, combined with the low cost of density functional theory, has promoted their broad unification in various atomistic electronic structure codes. By comparison, differences between QC algorithms tend to be more foundational. Unifying such algorithms within a single code therefore demands far-reaching flexibility of the code design. Compounded with the versatility required to treat the variety of target problems of interest, this renders the development of canonical QC simulation software a considerable challenge.

In this Paper, we introduce the open-source project QC Lab (Quantum--Classical Laboratory) as our contribution to addressing the aforementioned challenge. QC Lab is designed as a systematic and flexible infrastructure whose modular design supports cross-compatibility between QC algorithms on one hand and QC models on the other. It is intended to serve the development and application of emerging algorithms, without posing restrictions on models, allowing advances in QC dynamics to be leveraged for the full range of relevant target problems in chemistry, physics, and materials science. By decomposing algorithms and models into a series of tasks and ingredients that can be substituted and reused, it minimizes development efforts and code redundancy. We sought to make QC Lab as accessible as possible by constructing it as a Python package for easy integration in Python-based programs and notebooks. Even though flexibility and accessibility are first priority, QC Lab takes advantage of a number of design principles that allow it to also reach serviceable performance. Through this combination of factors, we envision QC Lab to meaningfully complement existing QC software, including Newton-X \cite{barbatti_--fly_2007}, SHARC \cite{richter_sharc_2011}, NEXMD \cite{malone_nexmd_2020}, pyUNIxMD \cite{lee_pyunixmd_2021}, JADE \cite{du_--fly_2015}, Hefei-NAMD \cite{zheng_ab_2019}, PYXAID \cite{akimov_pyxaid_2013, akimov_advanced_2014}, and Libra \cite{akimov_libra_2016, shakiba_libra_2022}. Our research group began developing QC Lab in late 2023, leading to an alpha build that was released in April 2025. Feedback solicited based on this build led to further improvements, which were implemented in the first stable build (QC Lab 1.0) that was just released under the Apache 2.0 License \cite{noauthor_github_2025}.

This Paper is intended to present the general ideas behind QC Lab. In the following sections, we will detail the overall organization, the modularity with regard to the development and combination of algorithms and models, and the current capabilities of QC Lab, after which we close with a brief outlook. For a comprehensive manual, we refer to the QC Lab online documentation \cite{noauthor_online_2025}. We note that, throughout this Paper, we are making deliberate use of capitalization of terms such as Algorithms and Models in reference to classes, objects, and functions in QC Lab, while casual use of such terminology is left uncapitalized.

\emph{Note specific to this preprint:} References are made to non-permanent web addresses in this document \cite{noauthor_github_2025, noauthor_online_2025}. These will be replaced by permanent repositories upon final publication.

\section{Package Organization}

\subsection{Python Package}

QC Lab ships as a Python package and is itself fully written in Python. There are various reasons why we chose to make QC Lab part of the Python ecosystem. First, it allows QC Lab to take advantage of the vast (and growing) supply of Python resources, including highly-optimized modules. In its current form, QC Lab is making use of NumPy \cite{harris_array_2020}, Numba (optional) \cite{lam_numba_2015}, h5py (optional) \cite{collette_h5pyh5py_2023}, mpi4py (optional) \cite{rogowski_mpi4pyfutures_2023}, as well as the built-in logging, multiprocessing, and functools modules, among others. Second, seeing the ever-increasing use of Python as a workhorse for scientific coding, releasing QC Lab as a Python package renders it straightforward to incorporate into many evolving coding projects, and to combine with an increasing suite of other scientific Python packages, including PySCF \cite{sun_pyscf_2018, sun_recent_2020} for electronic structure theory, HOOMD-blue \cite{anderson_hoomd-blue_2020} for molecular dynamics, and TenPy \cite{hauschild_efficient_2018} for tensor network states. Lastly, as a Python package, the downloading and installing of QC Lab proceeds fully automatically through the Python Package Index (PyPI) repository, facilitated by the pip package manager.

\begin{figure}[t!]
    \centering
    \includegraphics[width=.99\columnwidth]{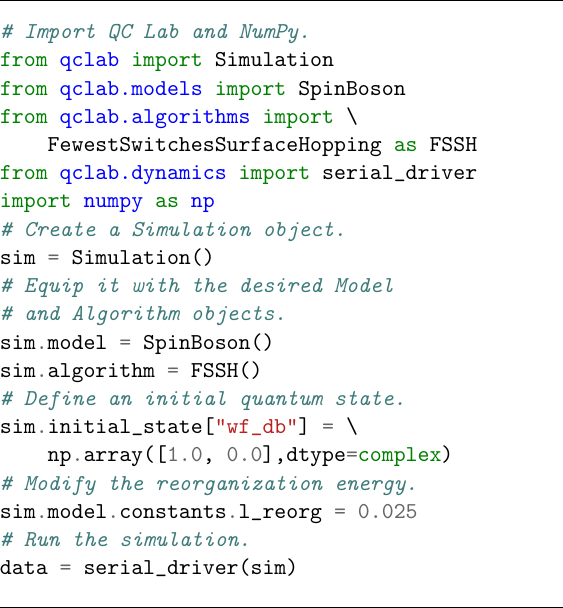}
    \caption{Example Python script for running an FSSH simulation of the spin--boson model using QC Lab. After importing the necessary modules, a Simulation object is created and equipped with the requisite Model and Algorithm objects and an initial quantum state. The reorganization energy is then changed from its default value before submitting the prepared Simulation object to the Dynamics Driver.}
    \label{fig:Spin-Boson}
\end{figure}

Once installed, QC Lab can be readily imported into a new or existing Python program or notebook; a workflow that sacrifices little to no conveniences that a stand-alone code would offer. The components of QC Lab are configured in such way that, once loaded, they come fully parametrized, with default parameter values typically correspond to a reference publication. As such, elementary simulations can be run with minimal effort. To demonstrate this, Fig.~\ref{fig:Spin-Boson} shows an example code for running a spin--boson model based on the fewest-switches surface hopping (FSSH) algorithm \cite{tully_molecular_1990}. Here, parameters are taken from Ref.~\citenum{tempelaar_generalization_2018}, while an adjustment is made to the reorganization energy.

\subsection{Overall Organization}\label{sec:overall}

\begin{figure}
    \centering
    \includegraphics[width=.99\columnwidth]{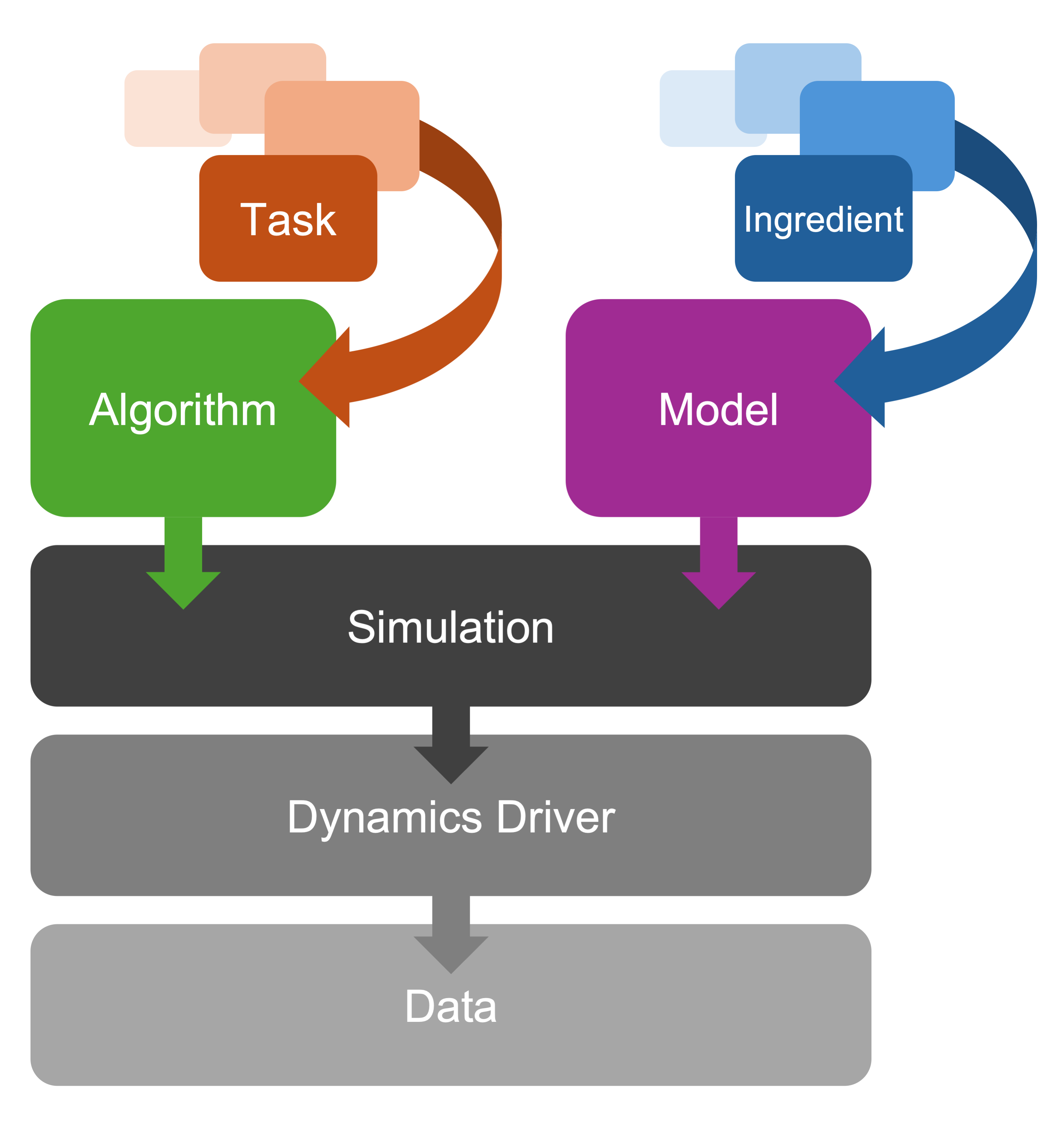}
    \caption{Schematic of the overall organization of QC Lab. Shapes depict objects or functions, and arrows indicate how such entities serve as input for other entities.}
    \label{fig:scheme}
\end{figure}

Shown in Fig.~\ref{fig:scheme} is a schematic of the overall organization of QC Lab. Fundamental to this organization is the reductionist notion that any QC simulation is composed of two coequal parts, namely a QC algorithm that is being employed, and a QC model problem which that algorithm is being applied to. During development of QC Lab we could not help but notice that a very similar breakdown applies to preparing a delicious meal, which consists of a list of consecutive tasks, referred to as a recipe. These tasks are applied to a list of ingredients. In designing the QC Lab organization, we have drawn from this analogy. Accordingly, central to QC Lab is an Algorithm object containing a chronological lists of Tasks, referred to as a Recipe, and a Model object containing a list of Ingredients.

\subsection{Simulation Object and Dynamics Driver}

The beating heart of QC Lab is the Dynamics Driver, which executes the QC simulation. The Dynamics Driver accepts a Simulation object that contains all information of a given simulation, including the Algorithm and Model objects, as well as simulation details that are not strictly tied to either a model or an algorithm. The latter includes the total runtime of a simulation and the state adopted for the initial quantum wavefunction. Running QC Lab entails calling the Dynamics Driver, which executes the QC simulation based on the Simulation object. At the outset of a simulation, the Dynamics Driver creates two Python dictionaries referred to as the State and Parameters objects. The State object contains all the transient quantities created during the execution of the QC algorithm. These include the quantum state and classical coordinates as well as intermediate quantities related to the algorithm's operations. The Parameters object similarly contains transient quantities, but instead is responsible for relaying them to the Model object where they can be utilized if the Model's properties depend on the state of the simulation. An example of this are time-dependent contributions to the quantum Hamiltonian, which are enabled by passing the current time of the simulation to the Model through the Parameters object.

\subsection{Data Object}

After completing the execution of the Algorithm, the Dynamics Driver returns a Data object. The Data object features a dictionary containing the output of the simulation, as well as a set of seeds used for any random number generations. The latter allows calculations to be reproduced at the single-trajectory level by relaying those seeds back to the Dynamics Driver. The Data object also contains a log storing information related to the operations of the simulation.

\subsection{Tasks and Ingredients}

\begin{figure}[h!]
    \centering
    \includegraphics[width=.99\columnwidth]{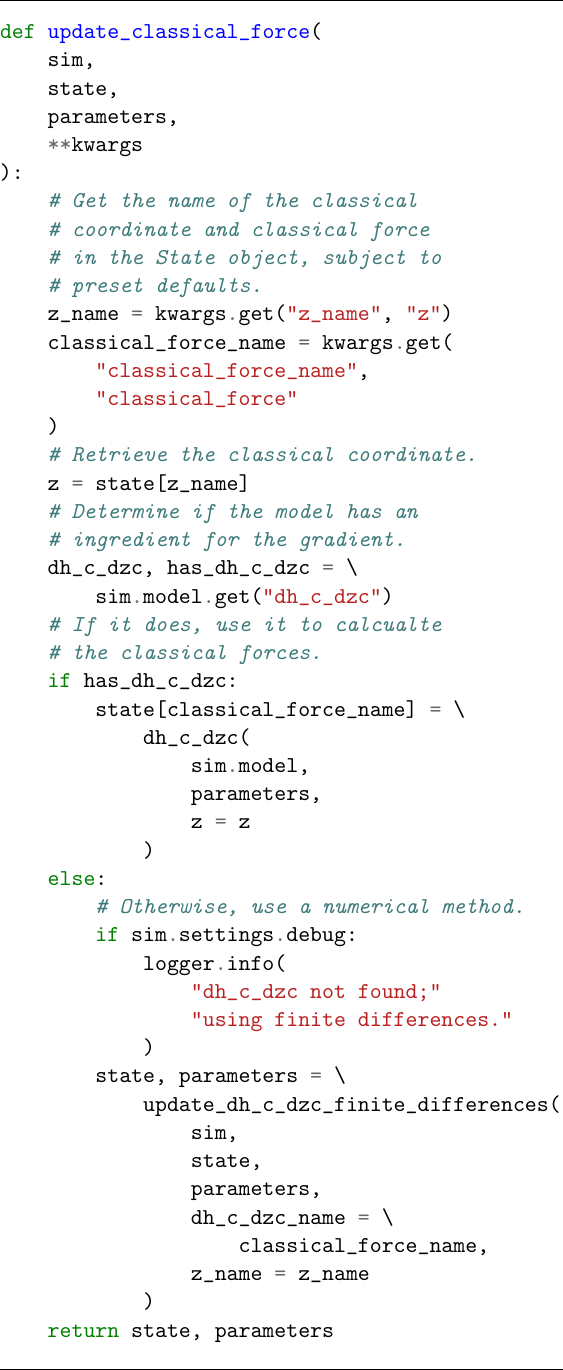}
    \caption{Example of a generic Task for calculating classical forces. Tasks follow a strict format for input and output arguments, but use of keyword arguments allows their action to be applied more generically. This example also shows the ability of Tasks to use Ingredients for optimized calculation of properties (in this case the gradient), which overrules default functionality.}
    \label{fig:Task}
\end{figure}

In QC Lab, Tasks are the building blocks of an Algorithm, and operate on the State and Parameters objects to update their contents. As mentioned previously, Tasks are organized within an Algorithm object by means of Recipes, which are lists that are executed in a consecutive order during a simulation. Tasks are Python functions that follow a strict and uniform format for the input and output arguments. This is illustrated in Fig.~\ref{fig:Task}, where we show the code of a generic Task responsible for calculating classical forces. As seen here, Tasks take as input the Simulation, State, and Parameters objects, whereas their output consists of the State and Parameters objects. The operation of Tasks is associated with specific contents of the State and Parameter objects. However, these associations can be overruled by use of keyword arguments, which are also accepted by the Tasks. This allows Tasks to perform their action as generically as possible, thereby helping to reduce code redundancy and increasing the transferrability of Tasks between different algorithms. As an example, the Task shown in Fig.~\ref{fig:Task} invokes the classical coordinates of the State object referred to as ``z'', but this association can be overruled by the keyword argument ``z\_name''.

On the other hand, Ingredients are the building blocks of a Model that generate the collection of physically-relevant quantities. At its most basic, this includes the Hamiltonians, but the collection of quantities can be expanded as needed. Like Tasks, Ingredients are Python functions following a strict input argument format. Specifically, Ingredients accept the Model and Parameters objects. Unlike Tasks, however, Ingredients follow an arbitrary output format, simply consisting of the quantities that they calculate. As a result, they need to be called from within the Tasks, while coordination of the output is established within the Task codes. The Ingredients are supplied to the Tasks via the Model object embedded in the Simulation object.

\section{Modularity}

\subsection{Algorithms and Models Cross Compatibility}

In designing QC Lab, we thought it critical that full cross compatibility is ensured between Algorithms and Models; not only those shipped with the current version, but also those to be added on in the coming years. This poses an interesting challenge since different algorithms require different pieces of information about the model to be able to operate. As an example, FSSH requires nonadiabatic coupling vectors, which are not required by Ehrenfest dynamics \cite{ehrenfest_bemerkung_1927, tully_mixed_1998}. Moreover, as-of-yet unanticipated algorithms may require further model details. How could we future proof QC Lab so that such algorithms will be compatible with the complete set of models?

In arriving at a solution, we took another reductionist approach, noting that a QC model is \emph{in principle} fully defined by means of a quantum Hamiltonian matrix, a classical Hamiltonian function, and a quantum--classical interaction Hamiltonian consisting of a matrix that parametrically depends on classical coordinates. QC Lab demands that, at a minimum, a Model comes with Ingredients specifying these three Hamiltonians. It further demands that each Algorithm object is able to fully function based entirely on these three Hamiltonians. The combination of both demands ensures cross-compatibility of any Algorithm with any Model, present and future.

If we had stopped there, QC Lab would have turned into an utterly unoptimized software. Indeed, depending on the algorithm, it would have to numerically calculate a series of derivative quantities based on the three Hamiltonians, examples of which include gradients with respect to classical coordinates necessary for nonadiabatic coupling vectors and quantum forces. While such quantities can be calculated based on finite-difference approaches, for many models there are analytical expressions available, the implementation of which would circumvent comparatively-expensive numerical calculations. However, those analytical expressions will be dependent on the model at hand, and may not be available in all cases. Those expressions can, therefore, not be incorporated as part of an Algorithm. On the other hand, we cannot demand them to be incorporated as a standard Ingredient in a Model either, because, by setting such standard, we may impede the implementation of future Algorithms that may need a \emph{different} set of model details, the composition of which we cannot currently foresee. Here, again, we run into an interesting challenge, as we are prompted with the question: How can we provide Algorithms with inexpensive analytical alternatives to numerical calculations without compromising cross-compatibility, present and future?

The solution we implemented in QC Lab is that Models may \emph{optionally} provide Ingredients to Tasks which will \emph{override} analogous functionalities within a given Task, while providing a performance boost based on analytical simplifications. Tasks involving computationally-heavy procedures are given the opportunity to check if such Ingredients are present in a given Model. If present, it will call those Ingredients, and if not present, it will resort to its own basic functionality instead. A concrete example of this is provided by the code shown in Fig.~\ref{fig:Task} for the Task responsible for calculating classical forces. Here, it is checked whether or not the applied Model provides an Ingredient supplying the gradient of the classical Hamiltonian with respect to the classical coordinates. If no such Ingredient is present, it uses a numerical method. By strictly enforcing such principles, we will ensure any future Algorithm to function with any Model, albeit not necessarily with high performance. Moreover, Models can be continually updated so as to provide performance-enhancing Ingredients for use by new Algorithms, without affecting their compatibility with any of the previous Algorithms.

\subsection{Modifying Algorithms and Models}

In addition to the cross compatibility between Algorithms and Models, the design of QC Lab is intended to streamline the process of developing \emph{new} Algorithms and Models. Overall, QC algorithms are largely built out of a set of common elements. Examples include a time-propagation routine for classical coordinates or the calculation of a quantum force acting on those coordinates. By dividing those elements into separate Tasks, a new Algorithm can be constructed by recycling existing Tasks where possible, and by only adding on new Tasks where the Algorithm differs from any of the pre-existing Algorithms. The same principles are applicable to Models, which tend to share common Ingredients. A new Model can thus be constructed by recycling existing Ingredients, while adding on Ingredients only where necessary. This not only minimizes the labor of implementing Algorithms and Models, but also avoids code redundancy.

\begin{figure}[t!]
    \centering
    \includegraphics[width=.99\columnwidth]{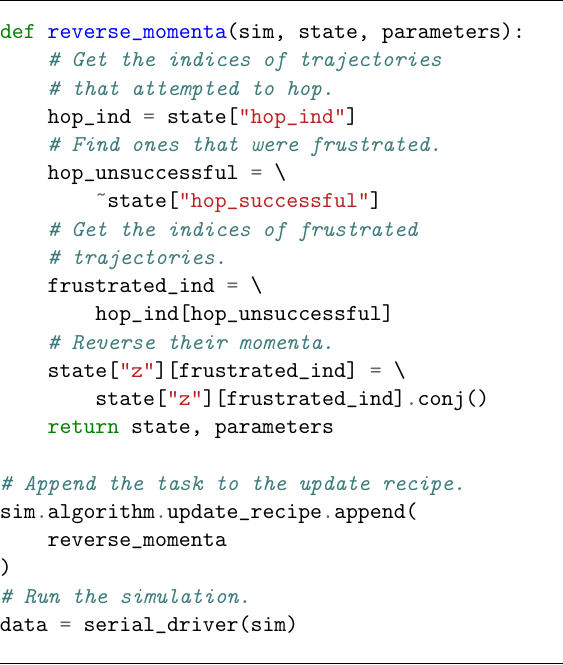}
    \caption{Modification of the simulation from Fig.~\ref{fig:Spin-Boson} such that momenta are reversed upon frustrated hops in the FSSH algorithm. After the simulation from Fig.~\ref{fig:Spin-Boson} is completed, running this code will adopt all Algorithm and Model settings from that simulation, while adding the Task ``reverse\_momenta'' to ``update\_recipe'' will invoke the momentum reversions. The latter is achieved by taking the complex conjugate of the complex classical coordinates, referred to as ``z''. We note that this procedure is only meaningful when the imaginary parts of these coordinates represents physical momenta, such as in a physical basis used for the spin--boson model. For a basis-independent procedure, a more involved approach is necessary.}
    \label{fig:Inverted}    
\end{figure}

The mixing and matching of Tasks and Ingredients into new Algorithms and Models is facilitated by lists contained in the Algorithm and Model objects, respectively. As mentioned in Sec.~\ref{sec:overall}, the Recipes appearing in the Algorithm object are chronological, as the Tasks in an Algorithm need to be performed in a consecutive order. The Model object, on the other hand, contains a list of Ingredients whose ordering is in principle irrelevant. It is, however, constructed in such way that every Ingredient is connected to a keyword that matches a call made by a Task in the Algorithm. Based on these principles, modifying Algorithms and Models amounts to adding necessary Tasks and Ingredients to the codebase, and then modifying the recipes and lists using standard Python commands in order to substitute those Tasks and Ingredients into the Algorithm and Model, respectively. Notably, the code for new Tasks and Ingredients can be added to dedicated modules in the QC Lab software, or to the Python program or notebook where QC Lab is imported (the latter option is generally preferred for casual use). Fig.~\ref{fig:Inverted} demonstrates a modification of the simulation from Fig.~\ref{fig:Spin-Boson} such that momenta are reversed upon frustrated hops in the FSSH algorithm.

\section{Capabilities}

\subsection{Representations From Real to Reciprocal Space}

It is typical for QC simulations to be implemented within a real-space basis (sometimes referred to as physical basis). However, in recent years there has been a growing interest in implementations that leverage alternative bases representations, including those adopted for the classical coordinates. For example, various studies have conducted QC simulations entirely in reciprocal space in order to efficiently capture the dynamics of crystalline solids \cite{krotz_reciprocal-space_2021, krotz_reciprocal-space_2022, xie_surface_2022, krotz_mixed_2024, wang_interband_2024, chen_floquet_2024, krotz_surface_2025}. Yet other basis representations have been explored in order to efficiently describe defective crystals \cite{miyazaki_mixed_2024}, drawing from the idea that QC dynamics can be reformulated by means of the full range of all possible unitary basis transformations \cite{miyazaki_mixed_2024}. Such basis representations are particularly relevant to the growing application of QC simulations to problems within the realms of condensed matter physics and materials science.

QC Lab is envisioned to facilitate the adoption of any arbitrary basis for the quantum states as well as the classical coordinates. This is realized by formulating Tasks in a way that is independent of any physical interpretation of the quantum and classical degrees of freedom, enabling such interpretation to be imposed after the fact and exclusively within the Model object. Ultimately, this enables the same physics to be calculated in different representations but with identical Algorithms. From the perception of the Tasks, the classical Hamiltonian simply assumes the form of a generic function of the classical coordinates. The quantum Hamiltonian, on the other hand, consists of a square matrix that generally is complex-valued, and which is provided in a diabatic (time-independent) basis commensurate with that of the vectors describing quantum states. The quantum--classical interaction Hamiltonian is another square and complex-valued matrix, expressed in the same diabatic basis, but which depends parametrically on the classical coordinates. By adequately configuring these functions and matrices in the Model object, the resultant simulation can be made to represent a problem in real space, reciprocal space, and beyond.

The lack of a predefined classical basis outside of the Model object comes with an important consequence that guided the incorporation of canonical coordinates in QC Lab. When adopting a physical basis for these coordinates, the quantum--classical interaction Hamiltonian depends parametrically only on the positions, and not on the momenta. However, once alternative basis representations are adopted, pairs of canonical coordinates may no longer correspond to physical positions and momenta. In such cases, a dependence on both coordinates in a pair, position \emph{and} momentum, may arise \cite{krotz_reciprocal-space_2021}. To facilitate such cases, and otherwise streamline the usage of canonical coordinates in QC Lab, we have combined each pair into a single complex-valued coordinate, such that position takes the real part, and momentum the imaginary part \cite{krotz_reciprocal-space_2021, miyazaki_mixed_2024}. Accordingly, QC Lab adopts a complex-valued formulation for classical dynamics \cite{miyazaki_mixed_2024}, but for convenience it is also equipped with standard functions that readily convert between this formulation and that based on pairs of positions and momenta.

We note that neither the basis set flexibility nor the adoption of complex-valued classical coordinates prevents the ability of QC Lab to support atomistic calculations. The atomic system could be incorporated within a Model object, or it could be outsourced to an existing molecular dynamics code by adapting the Algorithm object so that it extracts coordinates from that code rather than solving for the dynamical equations by itself. It would be particularly straightforward to combine QC Lab with HOOMD-blue \cite{anderson_hoomd-blue_2020}, as both take the form of Python packages that can be simultaneously imported in a program or notebook.

\subsection{Performance}

QC Lab intends to combine high flexibility with high performance. Performance optimization is primarily realized through the use of Python modules developed by others. This includes standard matrix diagonalization and multiplication tools from NumPy. In order to maximally harness the performance offered by such tools, QC Lab's core design vectorizes calculations by grouping them into batches of trajectories, thereby leveraging the trajectory-based nature of QC dynamics. Virtually all Tasks and Ingredients provided by QC Lab support such vectorization. However, to aid the development of Models by individuals less comfortable with vectorized programming, QC Lab also offers decorators for automatic vectorization of non-vectorized Ingredients. As a complementary resource, QC Lab is designed to be run through the Numba high-performance Python compiler, which conducts an on-the-fly translation of the Python language in QC Lab to optimized machine code.

QC Lab is equipped with various Dynamics Drivers designed to take advantage of different computing architectures and, in particular, the opportunities for parallelization they present. First, there is a serial Driver that lacks parallelization support, but instead is intended for casual use of QC Lab, for example when run at local desktop or laptop computers, or when debugging. Second, there is a Driver that harnesses Python's Multiprocessing module for parallelization. Lastly, there is a Driver that support parallelization through the Message Passing Interface (MPI) using the mpi4py module.

\subsection{Current Functionalities}

During the development of QC Lab, we have thus far prioritized delivering a flexible, robust, and polished infrastructure for QC simulations. We deliberately put less emphasis on shipping QC Lab with a comprehensive selection of Algorithms and Models. It should be noted that the addition of new Algorithms and Models is a relatively swift and easy process, owing to said infrastructure, and we expect such additions to be made by us and possibly by others in the near future.

The first stable release of QC Lab includes what can broadly be considered a minimal set of canonical Algorithms and Models. On the Algorithm side, it includes Ehrenfest dynamics (also known as mean-field dynamics) \cite{ehrenfest_bemerkung_1927, tully_mixed_1998} and FSSH \cite{tully_molecular_1990}. The latter is implemented in accordance with Ref.~\citenum{hammes-schiffer_proton_1994}, except that classical (physical) momenta are not reversed upon frustrated hops \cite{muller_surface-hopping_1997}. QC Lab is equipped with a few gauge fixing procedures that enforce parallel transport. Notably, for FSSH it offers flexibility to rescale classical momenta at will, allowing one to further impose specific gauge fixing constraints \cite{krotz_treating_2024}. Altogether, this allows FSSH to be applied to models involving geometric phase effects, including topological materials. On the Model side, QC Lab includes the spin--boson model, a one-electron Holstein lattice, and Tully's problems \cite{tully_molecular_1990}. In addition, QC Lab includes a Model representing the Fenna--Matthews--Olson light-harvesting complex, parametrized in accordance with Ref.~\citenum{mulvihill_simulating_2021}.

\section{Outlook}

As mentioned, this Paper marks the first stable release of QC Lab. In our anticipation, however, this release represents only a small fraction of the potential that QC Lab holds. By prioritizing its framework over its day-one capabilities, we have aimed to contribute a robust and lasting platform based on which the development and application of novel QC algorithms and models is facilitated. We intend to build an online infrastructure in order to turn such activities into a community effort. Much remains to be done in building this infrastructure, and in expanding the functionality of QC Lab. Regardless, we consider the current release an important stepping stone towards these future developments. In closing, we would like to refer once again to the documentation \cite{noauthor_online_2025}, which provides details well beyond this Paper, and which should offer the necessary guidance for interested individuals to get started with QC Lab. Our hope is that this software will serve novices and experts alike, and accelerate the development and adoption of QC dynamics simulations.

\section*{Acknowledgements}

The authors thank Anna Bondarenko for helpful discussions. This material is based upon work supported by the National Science Foundation under Grants No. 2145433 and No. 2513048. K.M. gratefully acknowledges support from the Mark A. Ratner Postdoctoral Fellowship and the Northwestern University International Institute for Nanotechnology (IIN).


\bibliography{qc_lab}

\providecommand{\latin}[1]{#1}
\makeatletter
\providecommand{\doi}
  {\begingroup\let\do\@makeother\dospecials
  \catcode`\{=1 \catcode`\}=2 \doi@aux}
\providecommand{\doi@aux}[1]{\endgroup\texttt{#1}}
\makeatother
\providecommand*\mcitethebibliography{\thebibliography}
\csname @ifundefined\endcsname{endmcitethebibliography}  {\let\endmcitethebibliography\endthebibliography}{}
\begin{mcitethebibliography}{43}
\providecommand*\natexlab[1]{#1}
\providecommand*\mciteSetBstSublistMode[1]{}
\providecommand*\mciteSetBstMaxWidthForm[2]{}
\providecommand*\mciteBstWouldAddEndPuncttrue
  {\def\EndOfBibitem{\unskip.}}
\providecommand*\mciteBstWouldAddEndPunctfalse
  {\let\EndOfBibitem\relax}
\providecommand*\mciteSetBstMidEndSepPunct[3]{}
\providecommand*\mciteSetBstSublistLabelBeginEnd[3]{}
\providecommand*\EndOfBibitem{}
\mciteSetBstSublistMode{f}
\mciteSetBstMaxWidthForm{subitem}{(\alph{mcitesubitemcount})}
\mciteSetBstSublistLabelBeginEnd
  {\mcitemaxwidthsubitemform\space}
  {\relax}
  {\relax}

\bibitem[Nelson \latin{et~al.}(2014)Nelson, Fernandez-Alberti, Roitberg, and Tretiak]{nelson_nonadiabatic_2014}
Nelson,~T.; Fernandez-Alberti,~S.; Roitberg,~A.~E.; Tretiak,~S. Nonadiabatic {Excited}-{State} {Molecular} {Dynamics}: {Modeling} {Photophysics} in {Organic} {Conjugated} {Materials}. \emph{Accounts of Chemical Research} \textbf{2014}, \emph{47}, 1155--1164\relax
\mciteBstWouldAddEndPuncttrue
\mciteSetBstMidEndSepPunct{\mcitedefaultmidpunct}
{\mcitedefaultendpunct}{\mcitedefaultseppunct}\relax
\EndOfBibitem
\bibitem[Subotnik \latin{et~al.}(2016)Subotnik, Jain, Landry, Petit, Ouyang, and Bellonzi]{subotnik_understanding_2016}
Subotnik,~J.~E.; Jain,~A.; Landry,~B.; Petit,~A.; Ouyang,~W.; Bellonzi,~N. Understanding the {Surface} {Hopping} {View} of {Electronic} {Transitions} and {Decoherence}. \emph{Annual Review of Physical Chemistry} \textbf{2016}, \emph{67}, 387--417\relax
\mciteBstWouldAddEndPuncttrue
\mciteSetBstMidEndSepPunct{\mcitedefaultmidpunct}
{\mcitedefaultendpunct}{\mcitedefaultseppunct}\relax
\EndOfBibitem
\bibitem[Wang \latin{et~al.}(2016)Wang, Akimov, and Prezhdo]{wang_recent_2016}
Wang,~L.; Akimov,~A.; Prezhdo,~O.~V. Recent {Progress} in {Surface} {Hopping}: 2011–2015. \emph{The Journal of Physical Chemistry Letters} \textbf{2016}, \emph{7}, 2100--2112\relax
\mciteBstWouldAddEndPuncttrue
\mciteSetBstMidEndSepPunct{\mcitedefaultmidpunct}
{\mcitedefaultendpunct}{\mcitedefaultseppunct}\relax
\EndOfBibitem
\bibitem[Crespo-Otero and Barbatti(2018)Crespo-Otero, and Barbatti]{crespo-otero_recent_2018}
Crespo-Otero,~R.; Barbatti,~M. Recent {Advances} and {Perspectives} on {Nonadiabatic} {Mixed} {Quantum}–{Classical} {Dynamics}. \emph{Chemical Reviews} \textbf{2018}, \emph{118}, 7026--7068\relax
\mciteBstWouldAddEndPuncttrue
\mciteSetBstMidEndSepPunct{\mcitedefaultmidpunct}
{\mcitedefaultendpunct}{\mcitedefaultseppunct}\relax
\EndOfBibitem
\bibitem[Herman and Kluk(1984)Herman, and Kluk]{herman_semiclasical_1984}
Herman,~M.~F.; Kluk,~E. A semiclasical justification for the use of non-spreading wavepackets in dynamics calculations. \emph{Chemical Physics} \textbf{1984}, \emph{91}, 27--34\relax
\mciteBstWouldAddEndPuncttrue
\mciteSetBstMidEndSepPunct{\mcitedefaultmidpunct}
{\mcitedefaultendpunct}{\mcitedefaultseppunct}\relax
\EndOfBibitem
\bibitem[Kay(2006)]{kay_hermankluk_2006}
Kay,~K.~G. The {Herman}–{Kluk} approximation: {Derivation} and semiclassical corrections. \emph{Chemical Physics} \textbf{2006}, \emph{322}, 3--12\relax
\mciteBstWouldAddEndPuncttrue
\mciteSetBstMidEndSepPunct{\mcitedefaultmidpunct}
{\mcitedefaultendpunct}{\mcitedefaultseppunct}\relax
\EndOfBibitem
\bibitem[Barbatti \latin{et~al.}(2007)Barbatti, Granucci, Persico, Ruckenbauer, Vazdar, Eckert-Maksić, and Lischka]{barbatti_--fly_2007}
Barbatti,~M.; Granucci,~G.; Persico,~M.; Ruckenbauer,~M.; Vazdar,~M.; Eckert-Maksić,~M.; Lischka,~H. The on-the-fly surface-hopping program system {Newton}-{X}: {Application} to ab initio simulation of the nonadiabatic photodynamics of benchmark systems. \emph{Journal of Photochemistry and Photobiology A: Chemistry} \textbf{2007}, \emph{190}, 228--240\relax
\mciteBstWouldAddEndPuncttrue
\mciteSetBstMidEndSepPunct{\mcitedefaultmidpunct}
{\mcitedefaultendpunct}{\mcitedefaultseppunct}\relax
\EndOfBibitem
\bibitem[Richter \latin{et~al.}(2011)Richter, Marquetand, González-Vázquez, Sola, and González]{richter_sharc_2011}
Richter,~M.; Marquetand,~P.; González-Vázquez,~J.; Sola,~I.; González,~L. {SHARC}: \textit{ab {Initio}} {Molecular} {Dynamics} with {Surface} {Hopping} in the {Adiabatic} {Representation} {Including} {Arbitrary} {Couplings}. \emph{Journal of Chemical Theory and Computation} \textbf{2011}, \emph{7}, 1253--1258\relax
\mciteBstWouldAddEndPuncttrue
\mciteSetBstMidEndSepPunct{\mcitedefaultmidpunct}
{\mcitedefaultendpunct}{\mcitedefaultseppunct}\relax
\EndOfBibitem
\bibitem[Malone \latin{et~al.}(2020)Malone, Nebgen, White, Zhang, Song, Bjorgaard, Sifain, Rodriguez-Hernandez, Freixas, Fernandez-Alberti, Roitberg, Nelson, and Tretiak]{malone_nexmd_2020}
Malone,~W.; Nebgen,~B.; White,~A.; Zhang,~Y.; Song,~H.; Bjorgaard,~J.~A.; Sifain,~A.~E.; Rodriguez-Hernandez,~B.; Freixas,~V.~M.; Fernandez-Alberti,~S.; Roitberg,~A.~E.; Nelson,~T.~R.; Tretiak,~S. {NEXMD} {Software} {Package} for {Nonadiabatic} {Excited} {State} {Molecular} {Dynamics} {Simulations}. \emph{Journal of Chemical Theory and Computation} \textbf{2020}, \emph{16}, 5771--5783\relax
\mciteBstWouldAddEndPuncttrue
\mciteSetBstMidEndSepPunct{\mcitedefaultmidpunct}
{\mcitedefaultendpunct}{\mcitedefaultseppunct}\relax
\EndOfBibitem
\bibitem[Lee \latin{et~al.}(2021)Lee, Ha, Han, Kim, Moon, and Min]{lee_pyunixmd_2021}
Lee,~I.~S.; Ha,~J.; Han,~D.; Kim,~T.~I.; Moon,~S.~W.; Min,~S.~K. {PyUNIxMD} : {A} \{{Python}‐based\} excited state molecular dynamics package. \emph{Journal of Computational Chemistry} \textbf{2021}, \emph{42}, 1755--1766\relax
\mciteBstWouldAddEndPuncttrue
\mciteSetBstMidEndSepPunct{\mcitedefaultmidpunct}
{\mcitedefaultendpunct}{\mcitedefaultseppunct}\relax
\EndOfBibitem
\bibitem[Du and Lan(2015)Du, and Lan]{du_--fly_2015}
Du,~L.; Lan,~Z. An {On}-the-{Fly} {Surface}-{Hopping} {Program} {JADE} for {Nonadiabatic} {Molecular} {Dynamics} of {Polyatomic} {Systems}: {Implementation} and {Applications}. \emph{Journal of Chemical Theory and Computation} \textbf{2015}, \emph{11}, 1360--1374\relax
\mciteBstWouldAddEndPuncttrue
\mciteSetBstMidEndSepPunct{\mcitedefaultmidpunct}
{\mcitedefaultendpunct}{\mcitedefaultseppunct}\relax
\EndOfBibitem
\bibitem[Zheng \latin{et~al.}(2019)Zheng, Chu, Zhao, Zhang, Guo, Wang, Jiang, and Zhao]{zheng_ab_2019}
Zheng,~Q.; Chu,~W.; Zhao,~C.; Zhang,~L.; Guo,~H.; Wang,~Y.; Jiang,~X.; Zhao,~J. Ab initio nonadiabatic molecular dynamics investigations on the excited carriers in condensed matter systems. \emph{WIREs Computational Molecular Science} \textbf{2019}, \emph{9}, e1411\relax
\mciteBstWouldAddEndPuncttrue
\mciteSetBstMidEndSepPunct{\mcitedefaultmidpunct}
{\mcitedefaultendpunct}{\mcitedefaultseppunct}\relax
\EndOfBibitem
\bibitem[Akimov and Prezhdo(2013)Akimov, and Prezhdo]{akimov_pyxaid_2013}
Akimov,~A.~V.; Prezhdo,~O.~V. The {PYXAID} {Program} for {Non}-{Adiabatic} {Molecular} {Dynamics} in {Condensed} {Matter} {Systems}. \emph{Journal of Chemical Theory and Computation} \textbf{2013}, \emph{9}, 4959--4972\relax
\mciteBstWouldAddEndPuncttrue
\mciteSetBstMidEndSepPunct{\mcitedefaultmidpunct}
{\mcitedefaultendpunct}{\mcitedefaultseppunct}\relax
\EndOfBibitem
\bibitem[Akimov and Prezhdo(2014)Akimov, and Prezhdo]{akimov_advanced_2014}
Akimov,~A.~V.; Prezhdo,~O.~V. Advanced {Capabilities} of the {PYXAID} {Program}: {Integration} {Schemes}, {Decoherence} {Effects}, {Multiexcitonic} {States}, and {Field}-{Matter} {Interaction}. \emph{Journal of Chemical Theory and Computation} \textbf{2014}, \emph{10}, 789--804\relax
\mciteBstWouldAddEndPuncttrue
\mciteSetBstMidEndSepPunct{\mcitedefaultmidpunct}
{\mcitedefaultendpunct}{\mcitedefaultseppunct}\relax
\EndOfBibitem
\bibitem[Akimov(2016)]{akimov_libra_2016}
Akimov,~A.~V. Libra: {An} open-{Source} “methodology discovery” library for quantum and classical dynamics simulations: {SOFTWARE} {NEWS} {AND} {UPDATES}. \emph{Journal of Computational Chemistry} \textbf{2016}, \emph{37}, 1626--1649\relax
\mciteBstWouldAddEndPuncttrue
\mciteSetBstMidEndSepPunct{\mcitedefaultmidpunct}
{\mcitedefaultendpunct}{\mcitedefaultseppunct}\relax
\EndOfBibitem
\bibitem[Shakiba \latin{et~al.}(2022)Shakiba, Smith, Li, Dutra, Jain, Sun, Garashchuk, and Akimov]{shakiba_libra_2022}
Shakiba,~M.; Smith,~B.; Li,~W.; Dutra,~M.; Jain,~A.; Sun,~X.; Garashchuk,~S.; Akimov,~A. Libra: {A} modular software library for quantum nonadiabatic dynamics. \emph{Software Impacts} \textbf{2022}, \emph{14}, 100445\relax
\mciteBstWouldAddEndPuncttrue
\mciteSetBstMidEndSepPunct{\mcitedefaultmidpunct}
{\mcitedefaultendpunct}{\mcitedefaultseppunct}\relax
\EndOfBibitem
\bibitem[noa(2025)]{noauthor_github_2025}
{GitHub} {Page}. 2025; \url{https://github.com/tempelaar-team/qclab}\relax
\mciteBstWouldAddEndPuncttrue
\mciteSetBstMidEndSepPunct{\mcitedefaultmidpunct}
{\mcitedefaultendpunct}{\mcitedefaultseppunct}\relax
\EndOfBibitem
\bibitem[noa(2025)]{noauthor_online_2025}
Online {Documentation}. 2025; \url{https://tempelaar-team.github.io/qclab/interactive_docs/index.html}\relax
\mciteBstWouldAddEndPuncttrue
\mciteSetBstMidEndSepPunct{\mcitedefaultmidpunct}
{\mcitedefaultendpunct}{\mcitedefaultseppunct}\relax
\EndOfBibitem
\bibitem[Harris \latin{et~al.}(2020)Harris, Millman, Van Der~Walt, Gommers, Virtanen, Cournapeau, Wieser, Taylor, Berg, Smith, Kern, Picus, Hoyer, Van~Kerkwijk, Brett, Haldane, Del~Río, Wiebe, Peterson, Gérard-Marchant, Sheppard, Reddy, Weckesser, Abbasi, Gohlke, and Oliphant]{harris_array_2020}
Harris,~C.~R.; Millman,~K.~J.; Van Der~Walt,~S.~J.; Gommers,~R.; Virtanen,~P.; Cournapeau,~D.; Wieser,~E.; Taylor,~J.; Berg,~S.; Smith,~N.~J.; Kern,~R.; Picus,~M.; Hoyer,~S.; Van~Kerkwijk,~M.~H.; Brett,~M.; Haldane,~A.; Del~Río,~J.~F.; Wiebe,~M.; Peterson,~P.; Gérard-Marchant,~P.; Sheppard,~K.; Reddy,~T.; Weckesser,~W.; Abbasi,~H.; Gohlke,~C.; Oliphant,~T.~E. Array programming with {NumPy}. \emph{Nature} \textbf{2020}, \emph{585}, 357--362\relax
\mciteBstWouldAddEndPuncttrue
\mciteSetBstMidEndSepPunct{\mcitedefaultmidpunct}
{\mcitedefaultendpunct}{\mcitedefaultseppunct}\relax
\EndOfBibitem
\bibitem[Lam \latin{et~al.}(2015)Lam, Pitrou, and Seibert]{lam_numba_2015}
Lam,~S.~K.; Pitrou,~A.; Seibert,~S. Numba: a {LLVM}-based {Python} {JIT} compiler. Proceedings of the {Second} {Workshop} on the {LLVM} {Compiler} {Infrastructure} in {HPC}. Austin Texas, 2015; pp 1--6\relax
\mciteBstWouldAddEndPuncttrue
\mciteSetBstMidEndSepPunct{\mcitedefaultmidpunct}
{\mcitedefaultendpunct}{\mcitedefaultseppunct}\relax
\EndOfBibitem
\bibitem[Collette \latin{et~al.}(2023)Collette, Kluyver, Caswell, Tocknell, Kieffer, Jelenak, Scopatz, Dale, {Chen}, VINCENT, Einhorn, {Payno}, {Juliagarriga}, {Pierlauro Sciarelli}, Valls, {Satrajit Ghosh}, Pedersen, Kittisopikul, {Jakirkham}, Raspaud, Danilevski, {Hameer Abbasi}, Readey, Mühlbauer, Paramonov, Chan, De~Schepper, {V. Armando Solé}, {Jialin}, and Guest]{collette_h5pyh5py_2023}
Collette,~A.; Kluyver,~T.; Caswell,~T.~A.; Tocknell,~J.; Kieffer,~J.; Jelenak,~A.; Scopatz,~A.; Dale,~D.; {Chen}; VINCENT,~T.; Einhorn,~M.; {Payno}; {Juliagarriga}; {Pierlauro Sciarelli}; Valls,~V.; {Satrajit Ghosh}; Pedersen,~U.~K.; Kittisopikul,~M.; {Jakirkham}; Raspaud,~M.; Danilevski,~C.; {Hameer Abbasi}; Readey,~J.; Mühlbauer,~K.; Paramonov,~A.; Chan,~L.; De~Schepper,~R.; {V. Armando Solé}; {Jialin}; Guest,~D.~H. h5py/h5py: 3.8.0-aarch64-wheels. 2023; \url{https://zenodo.org/record/7568214}\relax
\mciteBstWouldAddEndPuncttrue
\mciteSetBstMidEndSepPunct{\mcitedefaultmidpunct}
{\mcitedefaultendpunct}{\mcitedefaultseppunct}\relax
\EndOfBibitem
\bibitem[Rogowski \latin{et~al.}(2023)Rogowski, Aseeri, Keyes, and Dalcin]{rogowski_mpi4pyfutures_2023}
Rogowski,~M.; Aseeri,~S.; Keyes,~D.; Dalcin,~L. mpi4py.futures: {MPI}-{Based} {Asynchronous} {Task} {Execution} for {Python}. \emph{IEEE Transactions on Parallel and Distributed Systems} \textbf{2023}, \emph{34}, 611--622\relax
\mciteBstWouldAddEndPuncttrue
\mciteSetBstMidEndSepPunct{\mcitedefaultmidpunct}
{\mcitedefaultendpunct}{\mcitedefaultseppunct}\relax
\EndOfBibitem
\bibitem[Sun \latin{et~al.}(2018)Sun, Berkelbach, Blunt, Booth, Guo, Li, Liu, McClain, Sayfutyarova, Sharma, Wouters, and Chan]{sun_pyscf_2018}
Sun,~Q.; Berkelbach,~T.~C.; Blunt,~N.~S.; Booth,~G.~H.; Guo,~S.; Li,~Z.; Liu,~J.; McClain,~J.~D.; Sayfutyarova,~E.~R.; Sharma,~S.; Wouters,~S.; Chan,~G.~K. {PySCF}: the {Python}‐based simulations of chemistry framework. \emph{WIREs Computational Molecular Science} \textbf{2018}, \emph{8}, e1340\relax
\mciteBstWouldAddEndPuncttrue
\mciteSetBstMidEndSepPunct{\mcitedefaultmidpunct}
{\mcitedefaultendpunct}{\mcitedefaultseppunct}\relax
\EndOfBibitem
\bibitem[Sun \latin{et~al.}(2020)Sun, Zhang, Banerjee, Bao, Barbry, Blunt, Bogdanov, Booth, Chen, Cui, Eriksen, Gao, Guo, Hermann, Hermes, Koh, Koval, Lehtola, Li, Liu, Mardirossian, McClain, Motta, Mussard, Pham, Pulkin, Purwanto, Robinson, Ronca, Sayfutyarova, Scheurer, Schurkus, Smith, Sun, Sun, Upadhyay, Wagner, Wang, White, Whitfield, Williamson, Wouters, Yang, Yu, Zhu, Berkelbach, Sharma, Sokolov, and Chan]{sun_recent_2020}
Sun,~Q.; Zhang,~X.; Banerjee,~S.; Bao,~P.; Barbry,~M.; Blunt,~N.~S.; Bogdanov,~N.~A.; Booth,~G.~H.; Chen,~J.; Cui,~Z.-H.; Eriksen,~J.~J.; Gao,~Y.; Guo,~S.; Hermann,~J.; Hermes,~M.~R.; Koh,~K.; Koval,~P.; Lehtola,~S.; Li,~Z.; Liu,~J.; Mardirossian,~N.; McClain,~J.~D.; Motta,~M.; Mussard,~B.; Pham,~H.~Q.; Pulkin,~A.; Purwanto,~W.; Robinson,~P.~J.; Ronca,~E.; Sayfutyarova,~E.~R.; Scheurer,~M.; Schurkus,~H.~F.; Smith,~J. E.~T.; Sun,~C.; Sun,~S.-N.; Upadhyay,~S.; Wagner,~L.~K.; Wang,~X.; White,~A.; Whitfield,~J.~D.; Williamson,~M.~J.; Wouters,~S.; Yang,~J.; Yu,~J.~M.; Zhu,~T.; Berkelbach,~T.~C.; Sharma,~S.; Sokolov,~A.~Y.; Chan,~G. K.-L. Recent developments in the {PySCF} program package. \emph{The Journal of Chemical Physics} \textbf{2020}, \emph{153}, 024109\relax
\mciteBstWouldAddEndPuncttrue
\mciteSetBstMidEndSepPunct{\mcitedefaultmidpunct}
{\mcitedefaultendpunct}{\mcitedefaultseppunct}\relax
\EndOfBibitem
\bibitem[Anderson \latin{et~al.}(2020)Anderson, Glaser, and Glotzer]{anderson_hoomd-blue_2020}
Anderson,~J.~A.; Glaser,~J.; Glotzer,~S.~C. {HOOMD}-blue: {A} {Python} package for high-performance molecular dynamics and hard particle {Monte} {Carlo} simulations. \emph{Computational Materials Science} \textbf{2020}, \emph{173}, 109363\relax
\mciteBstWouldAddEndPuncttrue
\mciteSetBstMidEndSepPunct{\mcitedefaultmidpunct}
{\mcitedefaultendpunct}{\mcitedefaultseppunct}\relax
\EndOfBibitem
\bibitem[Hauschild and Pollmann(2018)Hauschild, and Pollmann]{hauschild_efficient_2018}
Hauschild,~J.; Pollmann,~F. Efficient numerical simulations with {Tensor} {Networks}: {Tensor} {Network} {Python} ({TeNPy}). \emph{SciPost Physics Lecture Notes} \textbf{2018}, 5\relax
\mciteBstWouldAddEndPuncttrue
\mciteSetBstMidEndSepPunct{\mcitedefaultmidpunct}
{\mcitedefaultendpunct}{\mcitedefaultseppunct}\relax
\EndOfBibitem
\bibitem[Tully(1990)]{tully_molecular_1990}
Tully,~J.~C. Molecular dynamics with electronic transitions. \emph{The Journal of Chemical Physics} \textbf{1990}, \emph{93}, 1061--1071\relax
\mciteBstWouldAddEndPuncttrue
\mciteSetBstMidEndSepPunct{\mcitedefaultmidpunct}
{\mcitedefaultendpunct}{\mcitedefaultseppunct}\relax
\EndOfBibitem
\bibitem[Tempelaar and Reichman(2018)Tempelaar, and Reichman]{tempelaar_generalization_2018}
Tempelaar,~R.; Reichman,~D.~R. Generalization of fewest-switches surface hopping for coherences. \emph{The Journal of Chemical Physics} \textbf{2018}, \emph{148}, 102309\relax
\mciteBstWouldAddEndPuncttrue
\mciteSetBstMidEndSepPunct{\mcitedefaultmidpunct}
{\mcitedefaultendpunct}{\mcitedefaultseppunct}\relax
\EndOfBibitem
\bibitem[Ehrenfest(1927)]{ehrenfest_bemerkung_1927}
Ehrenfest,~P. Bemerkung über die angenäherte {Gültigkeit} der klassischen {Mechanik} innerhalb der {Quantenmechanik}. \emph{Zeitschrift für Physik} \textbf{1927}, \emph{45}, 455--457\relax
\mciteBstWouldAddEndPuncttrue
\mciteSetBstMidEndSepPunct{\mcitedefaultmidpunct}
{\mcitedefaultendpunct}{\mcitedefaultseppunct}\relax
\EndOfBibitem
\bibitem[Tully(1998)]{tully_mixed_1998}
Tully,~J.~C. Mixed quantum–classical dynamics. \emph{Faraday Discussions} \textbf{1998}, \emph{110}, 407--419\relax
\mciteBstWouldAddEndPuncttrue
\mciteSetBstMidEndSepPunct{\mcitedefaultmidpunct}
{\mcitedefaultendpunct}{\mcitedefaultseppunct}\relax
\EndOfBibitem
\bibitem[Krotz \latin{et~al.}(2021)Krotz, Provazza, and Tempelaar]{krotz_reciprocal-space_2021}
Krotz,~A.; Provazza,~J.; Tempelaar,~R. A reciprocal-space formulation of mixed quantum–classical dynamics. \emph{The Journal of Chemical Physics} \textbf{2021}, \emph{154}, 224101\relax
\mciteBstWouldAddEndPuncttrue
\mciteSetBstMidEndSepPunct{\mcitedefaultmidpunct}
{\mcitedefaultendpunct}{\mcitedefaultseppunct}\relax
\EndOfBibitem
\bibitem[Krotz and Tempelaar(2022)Krotz, and Tempelaar]{krotz_reciprocal-space_2022}
Krotz,~A.; Tempelaar,~R. A reciprocal-space formulation of surface hopping. \emph{The Journal of Chemical Physics} \textbf{2022}, \emph{156}, 024105\relax
\mciteBstWouldAddEndPuncttrue
\mciteSetBstMidEndSepPunct{\mcitedefaultmidpunct}
{\mcitedefaultendpunct}{\mcitedefaultseppunct}\relax
\EndOfBibitem
\bibitem[Xie \latin{et~al.}(2022)Xie, Xu, Wang, and Zhuang]{xie_surface_2022}
Xie,~H.; Xu,~X.; Wang,~L.; Zhuang,~W. Surface hopping dynamics in periodic solid-state materials with a linear vibronic coupling model. \emph{The Journal of Chemical Physics} \textbf{2022}, \emph{156}, 154116\relax
\mciteBstWouldAddEndPuncttrue
\mciteSetBstMidEndSepPunct{\mcitedefaultmidpunct}
{\mcitedefaultendpunct}{\mcitedefaultseppunct}\relax
\EndOfBibitem
\bibitem[Krotz and Tempelaar(2024)Krotz, and Tempelaar]{krotz_mixed_2024}
Krotz,~A.; Tempelaar,~R. Mixed quantum–classical modeling of exciton–phonon scattering in solids: {Application} to optical linewidths of monolayer {MoS2}. \emph{The Journal of Chemical Physics} \textbf{2024}, \emph{161}, 044117\relax
\mciteBstWouldAddEndPuncttrue
\mciteSetBstMidEndSepPunct{\mcitedefaultmidpunct}
{\mcitedefaultendpunct}{\mcitedefaultseppunct}\relax
\EndOfBibitem
\bibitem[Wang and Dou(2024)Wang, and Dou]{wang_interband_2024}
Wang,~Y.; Dou,~W. Interband and intraband transitions, as well as charge mobility in driven two-band model with electron–phonon coupling. \emph{The Journal of Chemical Physics} \textbf{2024}, \emph{161}, 204104\relax
\mciteBstWouldAddEndPuncttrue
\mciteSetBstMidEndSepPunct{\mcitedefaultmidpunct}
{\mcitedefaultendpunct}{\mcitedefaultseppunct}\relax
\EndOfBibitem
\bibitem[Chen \latin{et~al.}(2024)Chen, Wang, and Dou]{chen_floquet_2024}
Chen,~J.; Wang,~Y.; Dou,~W. Floquet nonadiabatic mixed quantum–classical dynamics in periodically driven solid systems. \emph{The Journal of Chemical Physics} \textbf{2024}, \emph{160}, 214101\relax
\mciteBstWouldAddEndPuncttrue
\mciteSetBstMidEndSepPunct{\mcitedefaultmidpunct}
{\mcitedefaultendpunct}{\mcitedefaultseppunct}\relax
\EndOfBibitem
\bibitem[Krotz and Tempelaar(2025)Krotz, and Tempelaar]{krotz_surface_2025}
Krotz,~A.; Tempelaar,~R. Surface hopping simulations show valley depolarization driven by exciton-phonon resonance. 2025; \url{https://arxiv.org/abs/2505.06953}\relax
\mciteBstWouldAddEndPuncttrue
\mciteSetBstMidEndSepPunct{\mcitedefaultmidpunct}
{\mcitedefaultendpunct}{\mcitedefaultseppunct}\relax
\EndOfBibitem
\bibitem[Miyazaki \latin{et~al.}(2024)Miyazaki, Krotz, and Tempelaar]{miyazaki_mixed_2024}
Miyazaki,~K.; Krotz,~A.; Tempelaar,~R. Mixed {Quantum}–{Classical} {Dynamics} under {Arbitrary} {Unitary} {Basis} {Transformations}. \emph{Journal of Chemical Theory and Computation} \textbf{2024}, \emph{20}, 6500--6509\relax
\mciteBstWouldAddEndPuncttrue
\mciteSetBstMidEndSepPunct{\mcitedefaultmidpunct}
{\mcitedefaultendpunct}{\mcitedefaultseppunct}\relax
\EndOfBibitem
\bibitem[Hammes-Schiffer and Tully(1994)Hammes-Schiffer, and Tully]{hammes-schiffer_proton_1994}
Hammes-Schiffer,~S.; Tully,~J.~C. Proton transfer in solution: {Molecular} dynamics with quantum transitions. \emph{The Journal of Chemical Physics} \textbf{1994}, \emph{101}, 4657--4667\relax
\mciteBstWouldAddEndPuncttrue
\mciteSetBstMidEndSepPunct{\mcitedefaultmidpunct}
{\mcitedefaultendpunct}{\mcitedefaultseppunct}\relax
\EndOfBibitem
\bibitem[Müller and Stock(1997)Müller, and Stock]{muller_surface-hopping_1997}
Müller,~U.; Stock,~G. Surface-hopping modeling of photoinduced relaxation dynamics on coupled potential-energy surfaces. \emph{The Journal of Chemical Physics} \textbf{1997}, \emph{107}, 6230--6245\relax
\mciteBstWouldAddEndPuncttrue
\mciteSetBstMidEndSepPunct{\mcitedefaultmidpunct}
{\mcitedefaultendpunct}{\mcitedefaultseppunct}\relax
\EndOfBibitem
\bibitem[Krotz and Tempelaar(2024)Krotz, and Tempelaar]{krotz_treating_2024}
Krotz,~A.; Tempelaar,~R. Treating geometric phase effects in nonadiabatic dynamics. \emph{Physical Review A} \textbf{2024}, \emph{109}, 032210\relax
\mciteBstWouldAddEndPuncttrue
\mciteSetBstMidEndSepPunct{\mcitedefaultmidpunct}
{\mcitedefaultendpunct}{\mcitedefaultseppunct}\relax
\EndOfBibitem
\bibitem[Mulvihill \latin{et~al.}(2021)Mulvihill, Lenn, Gao, Schubert, Dunietz, and Geva]{mulvihill_simulating_2021}
Mulvihill,~E.; Lenn,~K.~M.; Gao,~X.; Schubert,~A.; Dunietz,~B.~D.; Geva,~E. Simulating energy transfer dynamics in the {Fenna}–{Matthews}–{Olson} complex via the modified generalized quantum master equation. \emph{The Journal of Chemical Physics} \textbf{2021}, \emph{154}, 204109\relax
\mciteBstWouldAddEndPuncttrue
\mciteSetBstMidEndSepPunct{\mcitedefaultmidpunct}
{\mcitedefaultendpunct}{\mcitedefaultseppunct}\relax
\EndOfBibitem
\end{mcitethebibliography}

\end{document}